\documentclass[letterpaper,american,prx,twocolumn,amsmath,amssymb,showpacs,superscriptaddress]{revtex4-1}
\usepackage[T1]{fontenc}
\usepackage{babel}
\usepackage{prettyref}
\usepackage{amstext}
\usepackage{graphicx}
\usepackage[unicode=true,
 bookmarks=true,bookmarksnumbered=false,bookmarksopen=false,
 breaklinks=false,pdfborder={0 0 1},backref=false,colorlinks=false]
 {hyperref}
\hypersetup{pdftitle={Spectral properties of excitable systems subject to colored noise},
 pdfauthor={Schuecker, Diesmann, Helias},
 pdfsubject={Manuscript for PRL},
 pdfkeywords={colored noise, integrate-and-fire neuron, transfer function},
 colorlinks=true,linkcolor=black,citecolor=black,urlcolor=black,filecolor=black}
\usepackage{breakurl}

\makeatletter

\newrefformat{cap}{\hyperref[#1]{Figure~\ref{#1}}}
\newrefformat{fig}{\hyperref[#1]{Figure~\ref{#1}}}
\newrefformat{tab}{\hyperref[#1]{Table ~\ref{#1}}}
\newrefformat{sec}{\hyperref[#1]{Section~\ref{#1}}}
\newrefformat{sub}{\hyperref[#1]{Section~\ref{#1}}}
\newrefformat{cha}{\hyperref[#1]{Chapter~\ref{#1}}}
\newrefformat{app}{\hyperref[#1]{Appendix~\ref{#1}}}

\makeatother

\begin{document}


\title{}

\title{A reaction diffusion-like formalism for plastic neural networks reveals
dissipative solitons at criticality}

\author{Dmytro Grytskyy }

\affiliation{Institute of Neuroscience and Medicine (INM-6) and Institute for
Advanced Simulation (IAS-6) }

\author{Markus Diesmann }

\affiliation{Institute of Neuroscience and Medicine (INM-6) and Institute for
Advanced Simulation (IAS-6) Department of Psychiatry, Psychotherapy
and Psychosomatics, Medical Faculty, RWTH Aachen University, Aachen,
Germany}

\affiliation{Department of Physics, Faculty 1, RWTH Aachen University, Aachen,
Germany}

\author{Moritz Helias}

\affiliation{Institute of Neuroscience and Medicine (INM-6) and Institute for
Advanced Simulation (IAS-6)}

\affiliation{Department of Physics, Faculty 1, RWTH Aachen University, Aachen,
Germany}

\date{\today}
\begin{abstract}
Self-organized structures in networks with spike-timing dependent
plasticity (STDP)  are likely to play a central role for information
processing in the brain. In the present study we derive a reaction-diffusion-like
formalism for plastic feed-forward networks of nonlinear rate neurons
with a correlation sensitive learning rule inspired by and being qualitatively
similar to STDP. After obtaining equations that describe the change
of the spatial shape of the signal from layer to layer, we derive
a criterion for the non-linearity necessary to obtain stable dynamics
for arbitrary input. We classify the possible scenarios of signal
evolution and find that close to the transition to the unstable regime
meta-stable solutions appear. The form of these dissipative solitons
is determined analytically and the evolution and interaction of several
such coexistent objects is investigated. 
\end{abstract}
\maketitle

\section*{Introduction}

Activity in neuronal networks influences their coupling structure
due to spike time-dependent synaptic plasticity (STDP) \citep{Bi98},
which, in turn, influences the activity. Several works analytically
investigated structures appearing in networks of neurons with plastic
synapses under the influence of external signals. These studies required
special network architectures, investigating e.g. a kind of possible
'elementary cell' of the network \citep{Nessler13_e1003037} or considered
all-to-all connected networks \citep{Galtier12} or averages over
activity realizations in space \citep{Takeuchi1979} or systems without
continuous spatial dimension \citep{Gilson09_1}.  Hebbian \citep{Hebb49}
and similar learning rules, which can be considered as a simple STDP-like
rules, are used in many neural network models, like e.g. Hopfield
and Boltzmann networks \citep{Chakrabarti08}. These works employ
the energy minimization principle known in physics and set the values
of synaptic weights according to the patterns to be stored, without
considering the time evolution of the weights explicitly. These systems
are able to store one level associations between patterns and recognize
and restore externally applied and previously learned patterns. The
memory capacity as well as times to recognize an input have been investigated
\citep{Chakrabarti08}.  Activity propagation in non-plastic feed-forward
systems was considered in \citep{Gajic13}. A set of integro-differential
equations, describing a spatially extended network, was introduced
in \citep{Wilson72_1} as the neural field model. A scheme of solution
for a simplified version neglecting refractoriness \citep{Amari77}
was later extended for neurons with adaptation \citep{Coombes05}
and different non-linearities \citep{Laing02}. The spatial and structural
organization of cortex \citep{Mountcastle97_701,Gonzalez00} separates
different neural inputs either in real space or in an effective space
that represents a continuum of features. For example, the orientation
selectivity of neurons in visual cortex is represented by neurons
topologically arranged on a one dimensional ring \citep{Ben-Yishai95}.
One suggested mechanism \citep{Compte00,Gutkin01} to implement short-term
memory is by localized bump solutions, which are also considered in
\citep{Laing02}. The latter work shows the relation between the formalism
of reaction-diffusion-like systems and spatially extended non-plastic
neural networks. Our work demonstrates such a relation for neural
networks with long term plasticity. The formalism thus fills the gap
between a series of studies of plastic networks without spatial dimension
and the formalism describing activity in spatially extended networks.
The presented analysis therefore opens the possibility to study the
transfer of short-term memory, encoded by bump solutions, into long-term
memory, stored by synaptic modifications. Here we analytically consider
a feed-forward network with space-dependent connectivity and linear-non-linear
neurons with a simple STDP-inspired synaptic learning rule, similar
to the BCM rule \citep{Bienenstock82}. We reduce the discrete problem
by diffusion approximation to obtain an equation similar but not exactly
equivalent to a reaction-diffusion equation with one component, also
called Kolmogorov-Petrovsky-Piskounov equation \citep{Kolmogorov37,Liehr13}.
We derive the requirements on the non-linearity necessary for a regime
of stable signal propagation, exposing that fine-tuning of parameters
is necessary, explaining earlier results \citep{Kunkel11_00160}.
The bump solutions that are stable in the critical and meta-stable
in the subcritical regime are described analytically. The interneural
connections inside a bump are strengthened, resembling cell assemblies
\citep{Vogels11_1569} and showing how externally presented objects
(external input) change intrinsic system properties (connectivity).

We further consider the interaction between bump solutions in dependence
of their size, the distance between them, and the system parameters.
In the regime close to stable propagation we find that several coexisting
activity bumps can either unite with each other or remain disjoint,
depending on the initial conditions. Their unification can be interpreted
as an emergence of connections between cell-assemblies, and, in this
way, represents a system of associations between internal representations
of corresponding external stimuli. The improved formalism can be generalized
to neural networks with several neuron types and to some extent mapped
onto the time evolution of a recurrent network, opening the possibility
for further investigations.

\section*{Results}

\subsection*{Model definition}

\begin{figure}
\begin{centering}
\includegraphics[angle=270,scale=0.7]{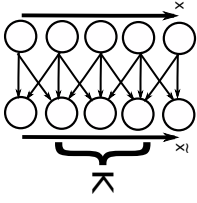}
\par\end{centering}

\caption{Two neighboring layers in the considered feed-forward architecture.
Synapses between layers obey the plasticity rule \eqref{eq:plasticity}.\label{fig:feed-forward}}
\end{figure}

Here we consider feed forward networks consisting of $M$ layers of
$N$ neurons each. A neuron at position $\tilde{x}$ in layer $n+1$
gets as input $u_{\tilde{x}}$ the weighted sum of outputs $r$ of
an odd number $K$ of neurons from the neighborhood in the previous
layer $n$, $u_{\tilde{x}}=\sum_{x=\tilde{x}-(K-1)/2}^{x=\tilde{x}+(K-1)/2}w_{\tilde{x}x}r_{x}$,
where the indices $\tilde{x}$ and $x$ describe the positions of
the postsynaptic neuron in the layer $n+1$, and of the presynaptic
neuron in the layer $n$, respectively. Here $w_{\tilde{x}x}$ is
the synaptic weight from neuron $x$ to neuron $\tilde{x}.$ The neuron's
output $r=f(u)$ is its input $u$ mapped by the non-linear function
$f$. This network architecture is illustrated in \prettyref{fig:feed-forward}.
The absence of feedback allows the solution of the system, finding
the activities and synaptic weights for each pair of layers separately
using the solution of the previous pair as an input.

We assume a Hebbian plasticity rule\vspace{-0.6cm}

\begin{equation}
\dot{w}_{\tilde{x}x}=-\alpha(w_{\tilde{x}x}-w_{0})+\gamma r_{x}r_{\tilde{x}}\label{eq:plasticity}
\end{equation}
for the presynaptic neuron $x$ and the postsynaptic neuron $\tilde{x}$,
which is similar to the BCM rule \citep{Bienenstock82}, but without
the ``sliding'' threshold in the non-linearity $f$. As a result,
the function $f$ keeps its sign. For simulations we used $f(x)=A/(1+e^{-\beta(x-\theta)})-A/(1+e^{\beta\theta})$.
If a stationary solution exists it can be found requiring $\dot{w}_{\tilde{x}x}=0$
as 
\begin{eqnarray}
w_{\tilde{x}x} & = & w_{0}+\frac{\gamma}{\alpha}r_{x}r_{\tilde{x}}=w_{0}+\frac{\gamma}{\alpha}f(u_{x})f(u{}_{\tilde{x}}).\label{eq:w_stationary}
\end{eqnarray}
In the following we use $\xi=u_{\tilde{x}}$ as the summed input to
the postsynaptic neuron to better distinguish it from the presynaptic
neuron's $u$.

In equilibrium, after the learning processes stopped, $\xi$ is to
be found as a solution of the self-consistent equation 
\begin{eqnarray}
\xi_{\tilde{x}} & = & \check{f}(r_{\xi})=\sum_{x=\tilde{x}-(K-1)/2}^{\tilde{x}+(K-1)/2}w_{\tilde{x}x}r_{x}\label{eq:self_cons_xi}\\
 & \stackrel{(\ref{eq:w_stationary})}{=} & w_{0}\sum_{x=\tilde{x}-(K-1)/2}^{\tilde{x}+(K-1)/2}r_{x}+\frac{\gamma}{\alpha}f(\xi)\sum_{x=\tilde{x}-(K-1)/2}^{\tilde{x}+(K-1)/2}r_{x}^{2},\nonumber 
\end{eqnarray}
with $r_{\xi}=f(\xi)$ and the symbol $\check{}$ denoting the
inverse function. In the approximation neglecting further derivatives
(and being exact for $K=3$ and $\partial_{x}$ denoting the discrete
lattice derivative) one can replace $\sum_{x=\tilde{x}-(K-1)/2}^{\tilde{x}+(K-1)/2}$
with $(K+a\partial_{x}^{2})$ where $a=\sum_{n=0}^{(K-1)/2}n^{2}=(K-1)K(K+1)/24$.
In doing so, we also make the transition from a discrete index $x$
to a continuous variable, also denoted as $x$. So, processes in the
system can be considered as an interplay between the diffusion described
by the $\partial_{x}^{2}$-operator and the explicit (given by $f$)
and implicit non-linearities, due to the term $r_{x}^{2}$ in \eqref{eq:self_cons_xi}
resulting from plasticity. The equation \eqref{eq:self_cons_xi} can
be rewritten in this approximation as

\begin{equation}
\xi=\check{f}(r_{\xi})=w_{0}(K+a\partial_{x}^{2})r+\frac{\gamma}{\alpha}r_{\xi}(K+a\partial_{x}^{2})r^{2}.\label{eq:diff_xi}
\end{equation}

\subsection*{Analysis of global stability and self-reproducing solutions}

\begin{figure}
\begin{centering}
\includegraphics[scale=0.5]{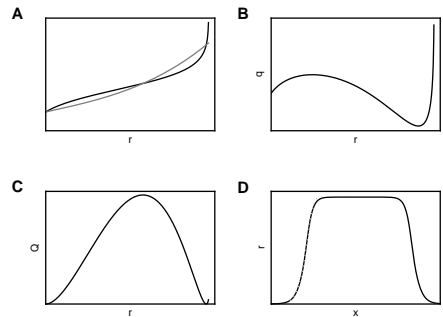}
\par\end{centering}

\caption{\textbf{A} Explicit $\check{f}(r)$ (black) and implicit $K(w_{0}+\frac{\gamma}{\alpha}r^{2})r$
non-linearity (gray). \textbf{B} Their difference, $q(r)$. \textbf{C}
$Q(r)=\int q(r)dr$, a minimum of $Q$ of $0$ enables a stable non-trivial
solution $r(x)$ shown in \textbf{D} - theoretical prediction of the
stable solution with dashed line and simulation results with solid
line. Simulation parameters: $N=800$, $M=400$, $K=41$, $w_{0}=0.99/K$,
$\gamma=0.99/K$, $\theta=0.6$, $\beta=3.6$, $A=1.0754$.\label{fig:stability_criteria}}
\end{figure}

One can search for possible stable solutions $\xi(x)=u(x)$, satisfying 

\begin{equation}
\check{f}(r)=w_{0}(K+a\partial_{x}^{2})r+\frac{\gamma}{\alpha}r(K+a\partial_{x}^{2})r^{2},\label{eq:diff_r_x}
\end{equation}
where $r(x)=f(u(x))$. There are no terms explicitly depending on
the variable $x$, so one can reduce the order of the equation by
introducing $y(r)=[\partial_{x}r](r)$ the derivative of $r$ depending
on $r$. We can hence express the second derivatives in \eqref{eq:diff_r_x}
as 
\begin{eqnarray}
\partial_{x}^{2}r & = & \partial_{x}y=\partial_{r}y\,\partial_{x}r=y\,\partial_{r}y\label{eq:del2r}\\
\partial_{x}^{2}(r^{2}) & = & 2(\partial_{x}r)^{2}+2r\,\partial_{x}^{2}r=2y^{2}+2ry\,\partial_{r}y=2z+r\,\partial_{r}z,\nonumber 
\end{eqnarray}
where we used the substitution $z(r)=y^{2}=(\partial_{x}r)^{2}$ in
the last step. Eq. \eqref{eq:diff_r_x} then takes the form of a linear
differential equation in $z$

\begin{eqnarray}
\check{f}(r) & = & w_{0}(K\,r+a\,y\,\partial_{r}y)+\frac{\gamma}{\alpha}(r^{3}K+ar\,(2z+r\,\partial_{r}z))\label{eq:diff_z_r}\\
 & = & w_{0}K\,r+\frac{\gamma}{\alpha}r^{3}K+2\frac{\gamma}{\alpha}ar\,z+(\frac{1}{2}w_{0}a+\frac{\gamma}{\alpha}ar^{2})\,\partial_{r}z,\nonumber 
\end{eqnarray}
replacing $y\,\partial_{r}y$ in the first line with $\frac{1}{2}\partial_{r}z$
in the second line. The solution of the corresponding homogeneous
equation $2\frac{\gamma}{\alpha}arz+(\frac{1}{2}w_{0}a+\frac{\gamma}{\alpha}ar^{2})\,\partial_{r}z=0$
is $H(r)=\exp(-\intop^{r}\frac{2\frac{\gamma}{\alpha}ar^{\prime}}{\frac{1}{2}w_{0}a+\frac{\gamma}{\alpha}ar^{\prime2}}\,dr^{\prime})=c\,\exp(-\ln(\frac{1}{2}w_{0}a+\frac{\gamma}{\alpha}ar^{2}))=c\frac{2}{a(w_{0}+2\frac{\gamma}{\alpha}r^{2})}$
(with an arbitrary constant $c$), and as solution of \eqref{eq:diff_z_r}
one gets

\begin{equation}
z(r)=\frac{2}{a(w_{0}+2\frac{\gamma}{\alpha}r^{2})}\int^{r}\underbrace{\left(\check{f}(r^{\prime})-K(w_{0}+\frac{\gamma}{\alpha}r^{\prime2})r^{\prime}\right)}_{\equiv q(r^{\prime})}\,dr^{\prime},\label{eq:sol_z_r}
\end{equation}
where we used that the factor in front of $\partial_{r}z$ cancels
with $H^{-1}$. As $z$ is the square of a real number and hence cannot
be negative, a solution existing for all $r$ (starting from $r=0$)
has to satisfy $Q(r)=\int_{0}^{r}q(r^{\prime})\,dr^{\prime}\geq0$.
We are seeking for solutions that start at $r=0$. We call $r_{\mathrm{max}}$
the largest value of $r$ reached, for which $\partial_{x}r=0$ and
therefore $z(r_{\mathrm{max}})=0$ from which follows

\begin{eqnarray}
Q(r_{\mathrm{max}}) & = & 0.\label{eq:stable_cond}
\end{eqnarray}
 For this case $Q(r)$, $q(r)$ and the stable solution $r(x)$ are
shown in \prettyref{fig:stability_criteria}. The physical meaning
of $q(r)$ is the competition between the non-linearity $f$ and the
effective non-linearity $K(w_{0}+\frac{\gamma}{\alpha}r^{2})r$ that
describe the propagation of $r$ neglecting diffusion: $K(w_{0}+\frac{\gamma}{\alpha}r^{2})r$
is the value of a postsynaptic neuron's membrane potential $\xi$
for the case that presynaptic and postsynaptic neurons have $u=\check{f}(r)$.
So, a positive $q(r)$ indicates a decrease of $u$ and $r$ from
one layer to the next, and negative $q(r)$ corresponds to both measures
increasing. So, in a stable system we must have $Q(r)\geq0$ for
all $r\geq0$. By definition, $Q(r)=0$ for $r=0$. If there are no
other $r$ at which $Q(r)=0$, there exists no self-consistent solution
propagating from layer to layer without change. If one tries to construct
the stable solution according to \eqref{eq:sol_z_r} for this case,
one will always have positive $\partial_{x}r$ for every $r>0$, which
means that for every $r>0$ 'external support' is needed to compensate
the diffusion effects, and without it every finite signal will decay
after some layers. For the case of the presence of an $r$ with $Q(r)=0$,
a solution exists having this $r$ as the maximum. If $q(r)=\partial_{r}Q(r)$
at that point is negative, an arbitrary small positive perturbation
added to this maximum $r$ will grow. So, $Q(r)<0$ for some $r$
means the absence of a stable solution and explosion of activity for
sufficiently strong activation patterns. 

The solution is given in the form $z(r)=(\partial_{x}r)^{2}$, where
the choice of a positive or negative sign in taking the square root
corresponds to a solution that is first increasing or decreasing,
respectively, when moving from left to right. The length of a ``plateau''
with $\partial_{x}r=0$ and $Q(r_{\mathrm{max}})=0$ is an arbitrary
parameter of the solution which can be chosen freely. If two such
solutions coexists, they lead to the growth of $r$ in the region
between borders of their plateaus and to their junction to one without
the external borders changing.

The form of the rest of the bump - the left and right wings $r=p(x)$
- is independent of the plateau's width and can be obtained from \eqref{eq:sol_z_r}
analytically or by numerical integration as the inverse function of
$\check{p}(r)=\int^{r}\pm1/\sqrt{z}\,dr=\int^{r}\pm1/\sqrt{\frac{2}{a(w_{0}+2\frac{\gamma}{\alpha}r^{2})}Q(r)}\,dr$
integrating from an arbitrary value of $r$ between $0$ and $r_{\mathrm{max}}$
and taking plus for the left or minus for the right wing.

\subsection*{Long-living solutions in sub-critical regime}

\begin{figure}
\begin{centering}
\includegraphics[scale=0.7]{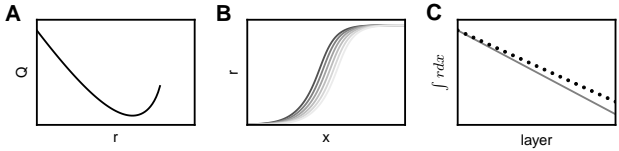}
\par\end{centering}

\caption{\textbf{A} $Q(r)$ near its minimum close to but bigger than $0$.
\textbf{B} Rate profiles $r(x)$ in different layers showing the reduction
of the plateau's width for a long-living rate profile leading to linear
decrease with the layer's number of $\int r(x)\,dx$. \textbf{C} The
theoretical expectation \eqref{eq:speed} and the simulated result
for $\int r(x)\,dx$. Parameters as for \prettyref{fig:stability_criteria}
apart from $A=1.0745$.\label{fig:speed_of_edge}}
\end{figure}

If no $r_{\mathrm{max}}>0$ with $Q(r_{\mathrm{max}})=0$ exists,
but $Q(\tilde{r}_{\mathrm{max}})$ has a local minimum close to zero,
one gets similar structures as the ``immortal'' solutions for \eqref{eq:stable_cond}.
The bump solutions propagate over large but finite numbers of layers
before they disappear. This number is approximately proportional to
the initial width $S$ of the plateau: for $S$ larger than $K$,
the successive reduction of this width does not affect the processes
on the borders, and the velocity $s$, i.e.the rate of reduction of
the plateau's width from layer to layer, is approximately constant,
depending only on network parameters as shown in \prettyref{fig:speed_of_edge}.
The velocity $s$ can be calculated as the propagation speed that
makes the solution stable. Obtaining the rate profile for the previous
layer as a slightly shifted version $r(x)-s\partial_{x}r(x)$ of
the profile $r(x)$ in the current layer one obtains from \eqref{eq:diff_xi}

\[
\check{f}(r)=w_{0}(K+a\partial_{x}^{2})(r-s\partial_{x}r)+\frac{\gamma}{\alpha}r(K+a\partial_{x}^{2})(r-s\partial_{x}r)^{2}
\]
or, after sorting, neglecting terms $O(s^{2})$, using the definition
of $q(r)$ in \eqref{eq:sol_z_r} and multiplying with $\partial_{x}r$

\begin{eqnarray*}
 &  & q(r)\partial_{x}r-sw_{0}K(\partial_{x}r)^{2}-sw_{0}a\partial_{x}^{3}r\,\partial_{x}r+aw_{0}\partial_{x}^{2}r\,\partial_{x}r=\\
\text{} &  & -2s\frac{\gamma}{\alpha}Kr^{2}(\partial_{x}r)^{2}+\frac{\gamma}{\alpha}ar\partial_{x}r\partial_{x}^{2}(r^{2})-2s\frac{\gamma}{\alpha}ar\partial_{x}^{2}(r\partial_{x}r)\partial_{x}r
\end{eqnarray*}
which, after integrating over $x$ from minus infinity to the beginning
of the plateau, leads to

\begin{eqnarray}
Q(\tilde{r}_{\mathrm{max}}) & = & -s\int_{0}^{\tilde{r}{}_{\mathrm{max}}}w_{0}K\partial_{x}r+aw_{0}\partial_{x}^{3}r+\frac{\gamma}{\alpha}(2Kr^{2}\partial_{x}r\nonumber \\
 & + & 2ar\partial_{x}^{2}(r\partial_{x}r)+ar\partial_{x}^{2}(r^{2}))+aw_{0}\partial_{x}^{2}r\,dr.
\end{eqnarray}
The last two terms in the latter expression vanish, because $\int_{0}^{\tilde{r}{}_{\mathrm{max}}}\partial_{x}^{2}r\ dr=\int_{0}^{\tilde{r}{}_{\mathrm{max}}}y\partial_{r}y\ dr=\frac{1}{2}(\partial_{x}r)^{2}|_{r=0}^{r=\tilde{r}{}_{\mathrm{max}}}=0$.
Further we have $\int_{0}^{\tilde{r}{}_{\mathrm{max}}}r\partial_{x}^{2}(r^{2})\ dr=\frac{1}{2}\int_{0}^{\tilde{r}{}_{\mathrm{max}}^{2}}\partial_{x}^{2}(r^{2})\ dr^{2}=\frac{1}{2}\int_{0}^{\tilde{r}^{2}{}_{\mathrm{max}}}y_{2}\partial_{r^{2}}y_{2}\ dr^{2}=\frac{1}{4}(\partial_{x}(r^{2}))^{2}|_{r=0}^{r=\tilde{r}{}_{\mathrm{max}}}=0$
with $y_{2}=[\partial_{x}(r^{2})](r^{2})$ similar to $y(r)=[\partial_{x}r](r)$,
because $\tilde{r}{}_{\mathrm{max}}$ is the plateau height and derivatives
of $r$-dependent functions are $0$. One can therefore obtain $s$
with

\begin{eqnarray}
s & = & -Q(\tilde{r}_{\mathrm{max}})/(\int_{0}^{\tilde{r}_{max}}(w_{0}K\partial_{x}r+2\frac{\gamma}{\alpha}Kr^{2}\partial_{x}r+\nonumber \\
 &  & 2\frac{\gamma}{\alpha}ar\partial_{x}^{2}(r\partial_{x}r)+aw_{0}\partial_{x}^{3}r)\,dr).\label{eq:speed}
\end{eqnarray}
One can replace $\partial_{x}r$ with the dependence $\partial_{x}r=\sqrt{z(r)}$
given by \eqref{eq:sol_z_r} obtained for $Q(\tilde{r}{}_{\mathrm{max}})=0$
- this approximation is meaningful for integrative quantities that
are not influenced strongly by the local variation of $Q$ near $\tilde{r}_{\mathrm{max}}$.
The chain rule allows the replacement of $\partial_{x}$ with $\partial_{x}r\partial_{r}$,
in this way one can express $s$  in terms of the integral of the
function containing $z(r)$ over $r$.

To understand how the system represents two or more coexistent signals,
we investigate the situation with two coexistent meta-stable solutions.
Two such bumps can unite into one if the minimal amplitude of $r$
between their borders becomes big enough for self-generated growth
before one of the plateaus disappears.

\begin{figure}
\begin{centering}
\includegraphics[scale=0.5]{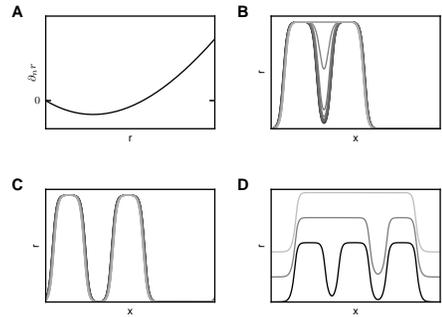}
\par\end{centering}

\caption{\textbf{A} Theoretical prediction of the change of minimal activity
$r_{\mathrm{min}}$ between bumps after one layer. \textbf{B} Propagation
of two bumps uniting to one, in \textbf{C} they remain disjoint, because
their initial distance exceeds the maximal distance for unification.
\textbf{D} Rate profiles in three layers. The activity of the bumps
is shifted upwards in proportion to the layer number for clearer distinction.
The evolution over layers illustrates the construction of an 'association
tree': first the left and the middle bumps are joint, and later the
right one. Parameters as for \prettyref{fig:speed_of_edge} apart
from $\gamma=0$, $w_{0}=1.4/K$, $\beta=3.63$ for panels \textbf{A}-\textbf{C}.\label{fig:multi_bumps}}
\end{figure}

Without loss of generality we take $x=0$ for the midpoint between
the bumps. If the distance between the two closer ends of the bumps'
plateaus is larger than some critical value, the minimal $r=r_{\mathrm{min}}$
at the minimum position $x=0$ decays. One can roughly estimate the
change $\delta r_{\mathrm{min}}$ of $r_{\mathrm{min}}$ from one
layer to the next, approximating $r(x)$ near $x=0$ with the direct
sum of the two bumps' wings, i.e. $r=p(x+\check{p}(r_{\mathrm{min}}/2))+p(-x+\check{p}(r_{\mathrm{min}}/2))$
with $p(x)$ denoting a wing of a bump with $p(0)=\tilde{r}_{\mathrm{max}}$
at the plateau's edge. The factor $1/2$ appears because the direct
sum of two identical bump solutions approximately produce the value
$r_{\mathrm{min}}$. In this approximation $\delta r_{\mathrm{min}}(r_{\mathrm{min}})=r_{\mathrm{min}}^{new}(r_{\mathrm{min}})-r_{\mathrm{min}}$,
where $r_{\mathrm{min}}^{new}=f([w_{0}(K+a\partial_{x}^{2})r+r(K+a\partial_{x}^{2})r^{2}]|_{x=0})$.
$\delta r_{\mathrm{min}}(r_{\mathrm{min}})$ is shown in \prettyref{fig:multi_bumps}.
The critical distance is given by $\check{p}(r_{\mathrm{min}}^{\mathrm{crit}}/2),\ \delta r_{\mathrm{min}}(r_{\mathrm{min}}^{\mathrm{crit}})=0$.
For larger distances no direct unification of two bumps is possible
independent of their plateau widths. 

The ``next generation'' of bumps resulting from this merging can
interact in the same way with each other and the bumps of previous
generations that are still alive. In this way an association tree
reflecting the input structure can be created as illustrated in \prettyref{fig:multi_bumps}.

\section*{Discussion}

We obtain equations describing the propagation of activity from layer
to layer in feed-forward networks of nonlinear rate neurons with an
STDP-inspired plasticity rule. We find that the stability of the considered
network is determined by the sign of the minimum of the function $Q(r)$:
for positive sign, every perturbation decays, for negative sign activity
explodes for any sufficiently strong perturbation. If the minimum
is $0$, stable attractive solutions exist. They have an analytically
obtained form containing a plateau of variable width. So, precise
tuning of parameters is required to get stable propagation of activity
patterns unchanged from layer to layer, in agreement with earlier
simulation results of \citep{Kunkel11_00160}. A sigmoidal form of
the nonlinearity $f$ can help to get an alteration of the signs of
$q=\partial_{r}Q(r)$, necessary to ensure the existence of a minimum
of $Q(r)=0$ at a non-vanishing activity level $r>0$. A sigmoidal
$f$ is therefore a natural choice to bring the system close to criticality.
 The qualitative form of the gain function found here is in line
with the sigmoidal form derived in \citep{Han99} for feed-forward
classification networks. The argumentation and the network model of
this earlier work are, however, quite different. Their model does
not include spatial organization. Rather, the authors find the sigmoid
resulting from the requirement of optimal stability in the recognition
of presented and previously learned patterns, without investigating
the learning process itself. For the latter they employ the known
error backpropagation mechanism, fixing the synaptic weights prior
to the consideration of the neuronal activation dynamics. Employing
field theoretic arguments, the sigmoid is found as a soliton (kink)
solution, where the input strength to the nonlinearity plays the role
of space. The dissipative soliton solutions in our work, in contrast,
are solutions in real space. Similar long-living bump solutions
exist in systems with positive minima of $Q(r)$ close to $0$. These
solutions decay with a velocity increasing with the value of this
minimum. Two bumps can unite if the interplay of diffusion and nonlinearity
between them overcomes the decay when propagating from one layer to
the next; otherwise they remain disjoint. For a united bump the same
scenarios exist, so a kind of association tree can appear in this
way. Qualitatively, all interesting results presented in this article
do not require the existence of the correlation sensitive term ($\gamma\neq0$)
and exist also in the system with static synapses, for which the correspondence
to differential equations is known \citep{Laing02}. For $\gamma=0$
\eqref{eq:diff_r_x} is rewritten as $q(r)=aw_{0}\partial_{x}^{2}r$,
describing the stable solution of a reaction-diffusion equation (RDE)
for one chemical with the reaction nonlinearity $q$ and diffusion
coefficient $aw_{0}$. One can also interpret the layers in the network
as different states of the system evolution in time on a discretized
time grid. The corresponding time-continuous equation $\partial_{t}u=aw_{0}\partial_{x}^{2}r-q(r)$
with redefined $q(r)=\check{f}(r^{\prime})/\tau_{m}-Kw_{0}r$ for
the presence of a linear leak $-u/\tau_{m}$ is analog to a RDE for
the time evolution of a system with one chemical. Dissipative solitons
are known to be solutions of such systems \citep{Liehr13} and correspond
to the bump solutions \eqref{eq:sol_z_r}. The interesting and non-trivial
result is, that such solutions exist also for arbitrary learning rates
$\gamma\neq0$ and that their shape can be obtained analytically.

A more general equation $q(r)=\sum_{i}D_{i}(r)\partial_{x}^{2}F_{i}(r)$
of the same type as \eqref{eq:diff_r_x}, but with an arbitrary number
of functions under the second derivative is solved with the same substitutions
$z,\ y$: $q(r)=\sum_{i}D_{i}(r)\partial_{x}^{2}F_{i}(r)=\sum_{i}D_{i}(r)((\partial_{x}^{2}r)\partial_{r}F_{i}(r)+(\partial_{x}r)^{2}\partial_{r}^{2}F_{i}(r))=\frac{1}{2}(\sum_{i}D_{i}(r)\partial_{r}F_{i}(r))\partial_{r}z+(\sum_{i}D_{i}(r)\partial_{r}^{2}F_{i}(r))z$,
$z(r)=2H(r)\int H^{-1}(r^{\prime})\frac{q(r^{\prime})}{\sum_{i}D_{i}(r^{\prime})\partial_{r^{\prime}}F_{i}(r^{\prime})}\ dr^{\prime}$
with $H(r)=\exp(-2\intop^{r}\frac{\sum_{i}D_{i}(r^{\prime})\partial_{r^{\prime}}^{2}F_{i}(r^{\prime})}{\sum_{i}D_{i}(r^{\prime})\partial_{r^{\prime}}F_{i}(r^{\prime})}\,dr^{\prime})$.
Further simplifications done for \eqref{eq:sol_z_r} were applicable
only to this particular problem and are impossible in the general
case.  This generalized equation can be used to describe e.g. stable
solutions for systems with several neuron and synapse types, in particular
networks including inhibitory neurons. There the interaction between
several bumps can exhibit more complex behavior than in the case of
non-plastic synapses \citep{Laing02}. This equation is similar but
not equivalent to an equation describing a one-component reaction-diffusion
system (for only one $i$  this mapping is exact).  Still, a similar
analysis of the existence of stable and meta-stable solutions, as
presented for \eqref{eq:diff_r_x}, is possible for this more general
case.  Preliminary results indicate that associative learning and
memory can persist even if the activity is restored to the baseline
level, similar as in \citep{Vogels11_1569}. In contrast to classical
models of associative memory, such as fully connected Hopfield networks
and Boltzmann machines \citep{Chakrabarti08}, our formalism allows
the study of spatially extended representations of earlier presented,
learned objects.

With the leak term $-u/\tau_{m}$, the evolution of activity is described
by $\partial_{t}u_{t}(\tilde{x})=-\tau_{m}^{-1}u_{t}(\tilde{x})+\sum_{x=\tilde{x}-(K-1)/2,\ x\neq\tilde{x}}^{\tilde{x}+(K-1)/2}w_{t}(\tilde{x},x)r_{t}(x),$
for stationary solutions $u(\tilde{x})=\sum_{x=\tilde{x}-(K-1)/2,\ x\neq\tilde{x}}^{\tilde{x}+(K-1)/2}\tau\,[w_{0}+\frac{\gamma}{\alpha}f(u(x))f(u(\tilde{x}))]f(u(x))$,
which, after diffusion approximation of the sum, is equivalent to
\eqref{eq:diff_r_x} with parameters $\tau\gamma$, $\tau w_{0}$
and $K-1$ instead of $\gamma$, $w_{0}$ and $K$. So the presented
results generalize to stationary solutions in recurrent networks.
\begin{quote}

\end{quote}

\end{document}